\documentclass[prl,twocolumn,showpacs]{revtex4}

\usepackage{graphicx}
\usepackage{amsmath}
\usepackage{amssymb}

\begin{document}
\def\a{\alpha}
\def\b{\beta}
\def\e{\varepsilon}
\def\d{\delta}
\def\l{\lambda}
\def\m{\mu}
\def\t{\tau}
\def\n{\nu}
\def\o{\omega}
\def\r{\rho}
\def\S{\Sigma}
\def\G{\Gamma}
\def\D{\Delta}
\def\O{\Omega}

\def\ra{\rightarrow}
\def\ua{\uparrow}
\def\da{\downarrow}
\def\pd{\partial}
\def\bg{{\bf g}}
\def\br{{\bf r}}
\def\bm{{\bf m}}
\def\bz{{\bf z}}

\def\be{\begin{equation}}
\def\ee{\end{equation}}
\def\bea{\begin{eqnarray}}
\def\eea{\end{eqnarray}}
\def\nn{\nonumber}
\def\lb{\label}
\def\pref#1{(\ref{#1})}

\title{Origin of four-fold anisotropy in square lattices of circular
ferromagnetic dots}

\author{G.N.~Kakazei,$^{1,2}$ Yu.G.~Pogorelov,$^3$ M.D.~Costa,$^4$ T.~Mewes,$^{5}$
P.E.~Wigen,$^2$ P.C.~Hammel,$^2$  V.O.~Golub,$^{1}$  T.~Okuno,$^6$
V.~Novosad$^7$}

\affiliation{$^1$Institute of Magnetism National Academy of Sciences
of Ukraine, 36b Vernadskogo Blvd., 03142 Kiev, Ukraine}

\affiliation{$^2$Department of Physics, Ohio State University, 191
West Woodruff Avenue, Columbus, OH 43210}

\affiliation{$^3$IFIMUP/Departamento de F\'{\i}sica, Universidade do
Porto, R. Campo Alegre, 687, Porto, 4169, Portugal}

\affiliation{$^4$CFP/Departmento de F\'{\i}sica, Universidade do
Porto, R. Campo Alegre, 687, Porto, 4169, Portugal}

\affiliation{$^5$MINT/Department of Physics and Astronomy,
University of Alabama, Box 870209, Tuscaloosa, AL 35487}

\affiliation{$^6$Institute for Chemical Research, Kyoto University,
Kyoto 611-0011, Japan}

\affiliation{$^7$Materials Science Division, Argonne National
Laboratory, Argonne, IL 60439}

\begin{abstract}
We discuss the four-fold anisotropy of in-plane ferromagnetic
resonance (FMR) field $H_r$, found in a square lattice of circular
Permalloy dots when the interdot distance $a$ gets comparable to the
dot diameter $d$. The minimum $H_r$, along the lattice
$\langle11\rangle$ axes, and the maximum, along the
$\langle10\rangle$ axes, differ by $\sim$ 50 Oe at $a/d$ = 1.1. This
anisotropy, not expected in uniformly magnetized dots, is explained
by a non-uniform magnetization $\bm(\br)$ in a dot in response to
dipolar forces in the patterned magnetic structure. It is well
described by an iterative solution of a continuous variational
procedure.
\end{abstract}
\pacs{75.10.Hk; 75.30.Gw; 75.70.Cn; 76.50.+g}
\maketitle

Magnetic nanostructures are of increasing interest for technological
applications, such as patterned recording media \cite{moser}, or
magnetic random access memories \cite{allwood}. One of the most
important issues for understanding their collective behavior is the
effect of long-range dipolar interactions between the dots
\cite{demokritov}. For the single-domain magnetic state of a dot,
the simplest approximation is that dots are uniformly magnetized and
interactions only define relative orientation of their magnetic
moments \cite{guslienko}. If so, the system of dipolar coupled dots
in a square lattice should be magnetically isotropic.

However, in all known experimental studies of closely packed arrays
of circular dots, a four-fold anisotropy (FFA) was found, either by
Brillouin light scattering \cite{mathieu}, ferromagnetic resonance
(FMR) \cite{jung} or magnetization measurements (from hysteresis
loops) \cite{natali,zhu}. It is important to note that FFA exists in
both unsaturated samples and saturated ones (i.e. above vortex
annihilation point on the hysteresis loop). Hence it cannot be only
associated with vortex formation suggested in Ref. \cite{natali}. It
was instead qualitatively related to stray fields from unsaturated
parts of magnetization inside the dots \cite{mathieu}. However no
quantitative description of FFA in such systems was given up to now.
So the aim of this study is to explain quantitatively the deviations
from isotropy in terms of modified demagnetizing effect in a
patterned planar system at decreasing inter-dot distance, from the
limit of isolated dot to that of continuous film. The choice of
\emph{X}-band FMR techniques for this study has an advantage in
eliminating possible interference from domain (vortex) structure
\cite{kakazei03}. The variational theoretical analysis is followed
by micromagnetic simulations.

Permalloy (Py) dots were fabricated with electron beam lithography
and lift-off techniques, as explained elsewhere \cite{novosad}. The
dots of thickness $t$ = 50 nm and diameter $d$ = 1 $\mu$m were
arranged into square arrays with the lattice parameter $a$ (center
to center distance) varying from 1.1 $\mu$m to 2.5 $\mu$m. The
dimensions were confirmed by atomic force microscopy and scanning
electron microscopy. Room temperature FMR studies were performed at
9.8 GHz using a standard \emph{X}-band spectrometer. The dependence
of the FMR field $H_r$ on the azimuthal angle $\varphi_H$ of applied
field $\mathbf{H}$ with respect to the lattice [10] axis for almost
uncoupled dots ($a$ = 2.5 $\mu$m) is shown in Fig. \ref{fig1}a. Only
a weak uniaxial anisotropy of $H_r(\varphi_H)$ is present here,
which can be fitted by the simple formula $H_{r}(\varphi_H) =
H_{r,av} + H_{2}\cos2\varphi_H$. For the $a$ = 2.5 $\mu$m sample, we
found the average peak position $H_{r,av} \approx 1.13$ kOe and the
uniaxial anisotropy field $H_{2} \approx 5$ Oe. The latter value
remains the same for the rest of our samples, so this uniaxial
anisotropy is most probably caused by some technological factors.

\begin{figure}
\includegraphics[bb=100bp 20bp 220bp 235bp, scale=.7]{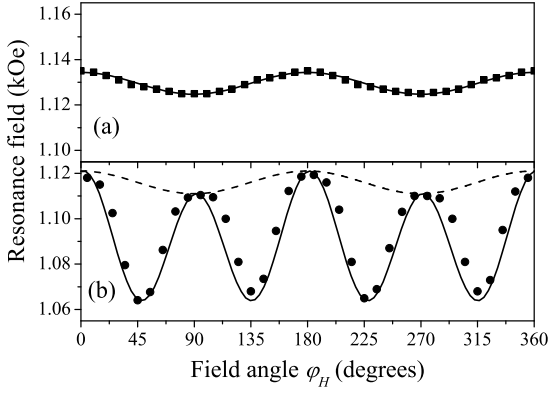}
\caption{In-plane FMR field in square lattices of 1 $\mu$m circular
Py dots as a function of field angle $\varphi_H$. a) The data for
lattice parameter $a = 2.5 \mu$m  are well fitted by uniaxial
anisotropy (solid line).  b) At $a = 1.1 \mu$m, the best fit (solid
line) is a superposition of FFA and uniaxial anisotropy (separately
shown by dashed line).} \lb{fig1}
\end{figure}

With decreasing distance $a$ between dots, two changes are observed
in the $H_r(\varphi_H)$ dependence. First, $H_{r,av}$ decreases to
$\approx 1.09$ kOe at $a$ = 1.1 $\mu$m (Fig. \ref{fig1}b). Second, a
four-fold anisotropy (FFA) is detected in the samples with
$a\leq$1.5 $\mu$m by pronounced minima of $H_{r}(\varphi_H)$ at
$\varphi_H$ close to the lattice $\langle11\rangle$ axes. This
behavior is fitted by $H_r(a,\varphi_H) = H_{r,av}(a) + H_4(a)
\cos4\varphi_H + H_2 \cos2\varphi_H$, as shown in Fig. \ref{fig1}b.
The interdot distance dependence of $H_{r,av}$ and FFA field $H_{4}$
is shown in Fig. \ref{fig2}. Also such anisotropy is detected in the
FMR linewidth, smaller for $\langle11\rangle$ than for
$\langle10\rangle$ case (reaching $\sim$30\% at $a/d$ = 1.1).

\begin{figure}
\includegraphics[bb=20bp 10bp 180bp 120bp, scale=1.3]{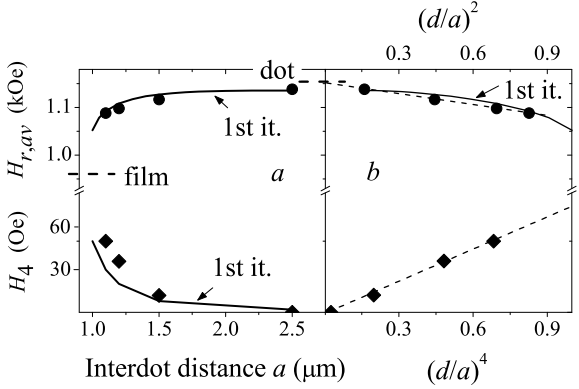}
\caption{ a) Average FMR field $H_{r,av}$ and FFA field $H_{4}$ as
functions of interdot spacing $a$. The points are the experimental
data and the solid lines present the 1st iteration theory (the
limits mark $H_r$ of isolated dot and continuous film). b) The same
data plotted against $(d/a)^2$ for $H_{r,av}$ and $(d/a)^4$ for
$H_{4}$ give excellent linear fits (dashed lines).} \lb{fig2}
\end{figure}

The FFA effect, which could not arise in uniformly in-plane
magnetized cylindrical dots, is evidently related to a non-uniform
distribution of the magnetization $\bm(r,\varphi,z)$ (in cylindric
coordinates $0 \leq r \leq R = d/2, 0 \leq \varphi < 2\pi, 0\leq z
\leq t$). A similar effect was discovered using Brillouin light
scattering \cite{mathieu} and magnetization reversal
\cite{natali,zhu} in such systems under weak enough external fields,
which displace vortices in each dot. This can be modeled by
displacements of two oppositely in-plane magnetized uniform domains
\cite{guslienko01}. But in the presence of external fields strong
enough to observe FMR, one has to assume a continuous (and mostly
slight) deformation of $\mathbf m(r,\varphi)$. The simplest model
for such deformation uses a variational procedure with respect to a
single parameter \cite{metlov}. However, as will be shown below, the
non-uniform magnetic ground state of this coupled periodic system
results from a rather complicate interplay between intra-dot and
inter-dot dipolar forces, which requires a more general variational
procedure.

Assuming fully planar and $z$-independent dot magnetization with the
2D Fourier amplitudes $\bm_\bg = \int \textrm{e}^{i\bg \cdot \br}
\bm(\br)d\br$, the total (Zeeman plus dipolar) magnetic energy (per
unit thickness of a dot) can be written as (see Appendix)

\be
\lb{eq1}
    E = -\mathbf{H} \cdot \bm_0 +
\frac{2 \pi} {a^2} \sum_{\bg\neq 0} \frac{f(g t)}{g^2} |\bm_\bg
\cdot \bg|^2, \ee

\noindent where $f(u)=1-\left(1-\textrm{e}^{-u}\right)/u$
\cite{guslienko} and the vectors of the 2D reciprocal lattice are
$\varphi_H$-dependent: $\bg = (2\pi/a) (n_1 \cos\varphi_H - n_2 \sin
\varphi_H, n_1 \sin \varphi_H + n_2 \cos \varphi_H)$ (for
$\mathbf{H}\parallel x$ and integer $n_{1,2}$). The variation of
exchange energy at deformations on the scale of whole sample is of
the order of stiffness constant ($\sim 10^{-6} \textrm{ erg/cm}$ for
Py) and it can be neglected beside the variation $\sim H M_s d^2
\sim 10^{-2} \textrm{ erg/cm}$ of terms included in Eq. \ref{eq1}.
If the dot magnetization has constant absolute value: $\bm (\br) =
M_s \left (\cos \varphi(\br), \sin \varphi(\br) \right)$, its
variation: $\d \bm (\br) = \hat{\bz} \times \bm (\br)
\d\varphi(\br)$ (where $\hat{\bz}$ is unit vector normal to plane),
is only due to the angle variation $\d \varphi (\br)$. Using the
Fourier transform $\d\bm_\bg = \hat{\mathbf{z}} \times
\sum_{\bg^\prime} \bm_{\bg - \bg^\prime}\d\varphi_{\bg^\prime}$ in
the condition $\d E = 0$ leads to the equilibrium equation for the
Fourier amplitudes:

\bea
 m_{\bg,y} &  = & \frac{4\pi}{H a^2} \sum_{\bg^\prime \neq 0}
\frac{f(g^\prime t)} {{g^\prime}^2}
(\bm_{\bg^\prime} \cdot \bg^\prime)   \notag  \\
& & \qquad\qquad\qquad (\bm_{\bg - \bg^\prime} \times \bg^\prime)
\cdot \hat{\bz}. \lb{eq2} \eea

\noindent It can be suitably solved by iterations:

\bea
 m^{(n)}_{\bg,y} &=& \frac{4\pi}{H
a^2}\sum_{\mathbf{g}^\prime\neq 0} \frac{f(g^\prime
t)}{{g^\prime}^2} (\bm^{(n-1)}_{\bg^\prime} \cdot
\bg^\prime)\nn \\
& & \,\qquad\qquad (\bm^{(n-1)}_{\bg - \bg^\prime}
\times\bg^\prime)\cdot\hat{\bz}, \lb{eq3} \eea

\noindent starting from uniformly magnetized dots as zeroth
iteration: $m^{(0)}_{\bg,y} = 0$, $m^{(0)}_{\bg,x} = 2\pi R M_s
J_1(g R)/g$  (with the Bessel function $J_1$). Already the 1st
iteration (including the inverse Fourier transform):

\bea
{m}^{(1)}_y(\br)& =& -\theta(d-2r) \frac{8\pi^2 R^2 M^2_s}{H
a^2} \sum_{\bg \neq 0}  \frac{f(g t) g_x g_y}{g^3}\,\nn \\
& &\qquad\qquad\qquad\qquad J_1(g R) \cos(\bg\cdot\br), \lb{eq5}
\eea

\noindent (with the Heavyside $\theta$ function) reveals the FFA
behavior, due to the rotationally non-invariant product $g_x g_y $.
The calculated maximum variation of $\varphi(\br) = \arcsin [m_y
(\br)/M_s]$ in $\langle10\rangle$  field geometry is $\sim60\%$
bigger than in the $\langle11\rangle$  geometry (Fig. \ref{fig3},
upper row). This expected behavior persists upon further iterations.
Our analytic approach was checked, using the micromagnetic OOMMF
code \cite{donahue} on a 9$\times$9 array of considered disks (Fig.
\ref{fig3}, lower row) at standard values of $M_{s} = 0.83$ kOe and
exchange stiffness $1.3 \cdot 10^{-6}$ erg/cm \cite{kakazei04} for
Py. The distributions obtained in this way for the central disk in
the array are within $\sim10\%$ to the analytic results of the 1st
iteration.

\begin{figure}
\includegraphics[width=8cm]{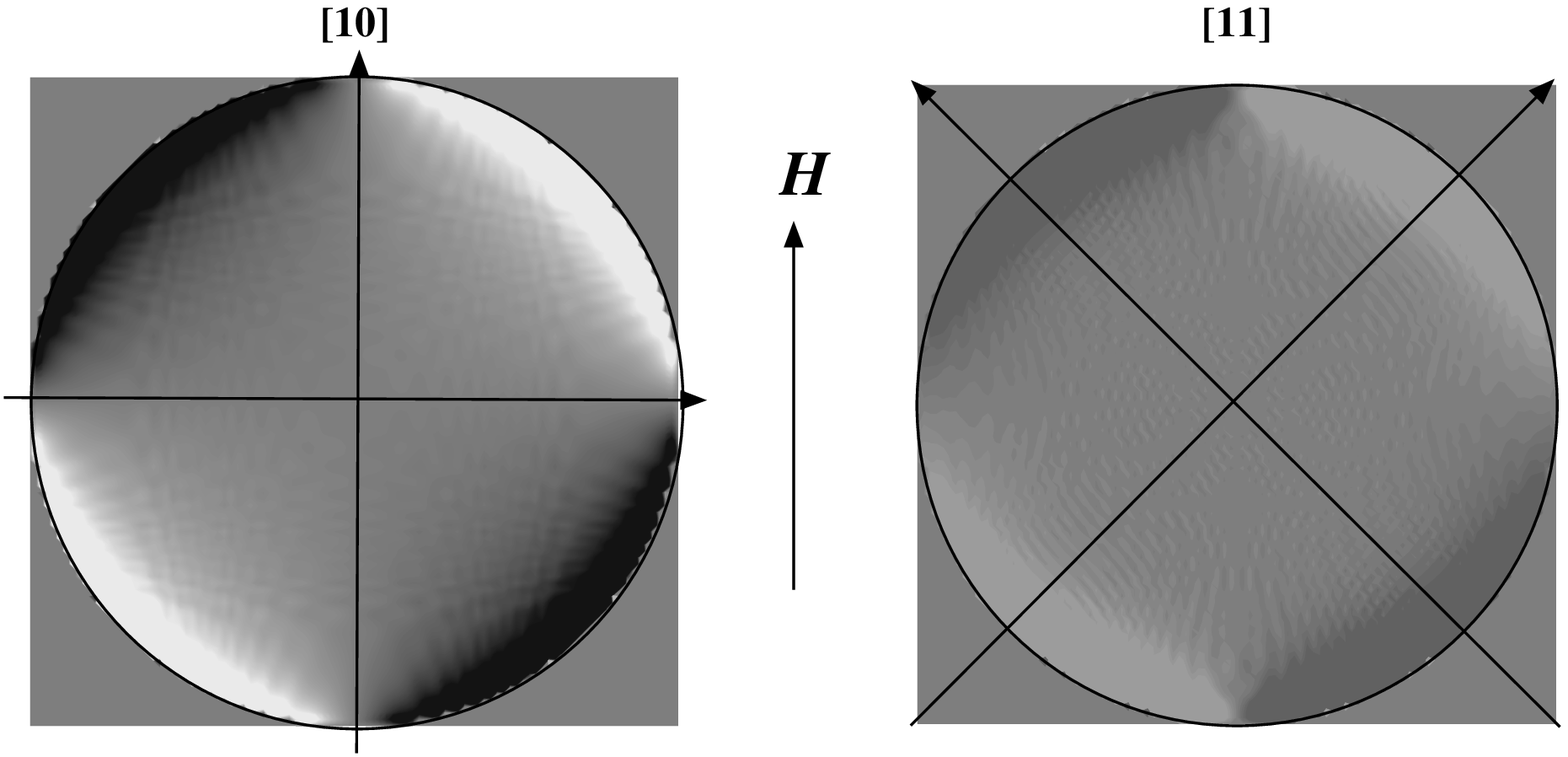}
\end{figure}
\begin{figure}
\includegraphics[width=8cm]{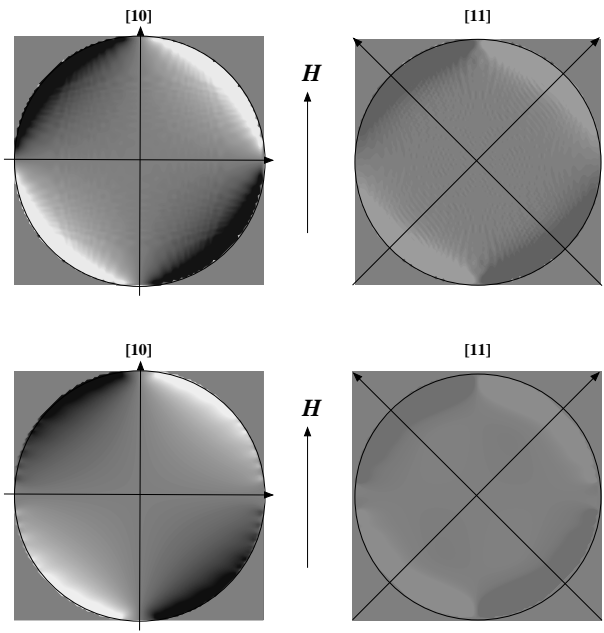}
\caption{ Density plots of the equilibrium magnetization angle
$\varphi(\mathbf r)$ for two field geometries, $\varphi_H = 0$
($\langle10\rangle$) and $\varphi_H = \pi/4$ ($\langle11\rangle$).
Upper row: calculated from the sum, Eq. \ref{eq5}, over
100$\times$100 sites of reciprocal lattice at parameter values $a =
1.1\, \mu$m, $H=$ 1.1 kOe, $M_s=$ 0.83 kOe. Lower row: micromagnetic
calculation by OOMMF code for the central disc in the 9$\times$9
array.} \lb{fig3}
\end{figure}

The FMR precession of $\bm(\br)$ is defined by the internal field
$\mathbf{H}_i(\br) = \mathbf{H} + \mathbf{h}(\br)$ through the local
dipolar field

\bea h_z(\br)&=&-\frac{4\pi}{a^2}\sum_{\bg}\left[1-f(g
t)\right]m_{\bg,z}\cos\left(\bg\cdot\br\right),\nn\\
h_\a(\br)&=&-\frac{4\pi}{a^2}\sum_{\b,\bg\neq 0} \frac{f(g t)g_\a
g_\b}{g^2}\tilde{m}_{\bg,\b}\cos\left(\bg\cdot\br\right), \lb{eq6}
\eea

\noindent ($\a,\b=x,y$). The 1st iteration for $\mathbf{h}(\br)$
corresponds to the zeroth iteration for $\bm(\br) =
(M_s,\m_y,\m_z)$, which now includes the uniform FMR amplitudes
$\m_y,\,\m_z$. Then the local demagnetizing factors $N_x(\br) = -
h_x(\br) / M_s,\,N_{y,z}(\br) = -h_{y,z}(\br)/\m_{y,z}$ define the
local FMR field $H_r(\br)$:

\bea
H_r(\br) &=& \sqrt{H_0^2+M_s^2[N_z(\br)-N_y(\br)]^2/4}\nn \\
   &-& M_s[N_z(\br)+N_y(\br)-2N_x(\br)]/2\lb{eq7}
\eea

\noindent (here $H_{0}\approx 3.3$ kOe). The average FMR field is
defined by the isotropic averaged demagnetizing factors

\begin{eqnarray}
\overline{N}_{x,y}& = &\left(\pi R^2 \right)^{-1} \int_{r < R}
N_{x,y}
\left(\br\right)d\br \nonumber\\
&=& (8\pi^2/a^2)\sum_{\bg\neq 0}f(g t)J_1^2(g R)/g^2, \label{eq8}
\end{eqnarray}

\noindent and $\overline{N}_z = 4\pi-2\overline{N}_x$. At
$a\to\infty$, they tend to the single dot values \cite{joseph} which
are for $t/R=0.1$: $N_{x,y}^{(d)}\approx0.776$ and $N_z^{(d)}
\approx 11.01$. Using $N_i^{(d)}$ instead of $N_i(\br)$ in Eq.
\ref{eq7} accurately reproduces the single dot FMR limit
$H_{r}^{(d)} \approx 1.15$ kOe (estimated from Fig. \ref{fig2}b).
Otherwise, for decreasing interdot distance, $a\to d$, the 1st
iteration values, Eq. \ref{eq8}, used in Eq. \ref{eq7} well describe
the tendency of $H_{r,av}(a)$ towards the continuous film limit
$H_r^{(f)}=\sqrt{H_0^2+4\pi^2M_s^2}-2\pi M_s\approx0.96$ kOe (Fig.
\ref{fig2}a).

Finally, by calculating the true local FMR fields $H_{r}(\br)$ from
Eqs. \ref{eq6} and \ref{eq7}, the field dependent absorption is
obtained as $I(H)\propto\int_{r<R} \d(H-H_{r}(\br))d\br$. Then the
FMR fields, $H_r$ defined from maximum of $I(H)$ in two geometries,
display FFA in a good agreement with the experimental data (Fig.
\ref{fig2}). This effect is due to the fact that stronger
deformation of magnetization stronger suppresses the demagnetizing
effect (the differences $N_z-N_{x,y}$) and thus enhances $H_r$. Also
it produces a bigger spread of local resonance fields $H_r(\br)$ and
thus broadens the FMR line, again in agreement with our
observations.

In conclusion, it is shown that under in-plane magnetic fields,
$\mathbf{H}$, even strong enough for FMR, the dipolar coupling in a
dense lattice of circular magnetic dots is able to produce a
continuous deformation of the dot magnetization, strongest for the
field orientation along lattice axes.

Work at ANL was supported by the U.S. Department of Energy, BES
Materials Sciences under Contract No. W-31-109-ENG-38; MDC was
supported by FCT (Portugal) and the European Union, through POCTI
(QCA III) grant No. SFRH/BD/7003/2001.

\section{Appendix}
For fully planar and \emph{z}-independent dot magnetization, the
dipolar energy per unit thickness of a dot in the lattice is:

 \bea
 E_d & = & \frac 1 {2t} \int_{-t/2}^{t/2} dz \int_{-t/2}^{t/2}
dz^\prime \int_c d\br \int d\br^\prime \sum_{\a,\b} m_\a(\br)\nn\\
& \times& \frac{\pd^2}{\pd r_\a \pd r_\b}\frac {m_\b(\br^\prime)}
{\sqrt{\left|\br - \br^\prime\right|^2 + \left(z -
z^\prime\right)^2}},
 \nn
 \eea

\noindent where the 2D integrations $\int_c d\br$ and $\int d\br$
are respectively over the unit cell and over the entire plane. It
can be also presented as

\bea E_d & = & \frac 1 {2t} \int_{-t/2}^{t/2} dz \int_c d\br
\sum_{\a} m_\a(\br) h_\a(\br,z) \nn\\
& =& \frac 1 {4\pi t a^2} \int_{-t/2}^{t/2} dz
\int_{-\infty}^{\infty} dq {\rm e}^{-i q z} \sum_{\a,\bg} \tilde
m_{\a,\bg} \tilde h_{\a,\bg,q},
 \nn
 \eea

\noindent where the Fourier amplitudes of the dipolar field are:

 \bea
  h_{\a,\bg,q} & = & \int_c d\br \int_{-\infty}^{\infty} dz^\prime
 {\rm e}^{i (\bg \cdot \br + qz^\prime)} h_\a(\br,z^\prime) \nn\\
 & = & \int_c d\br \int_{-\infty}^{\infty} dz^\prime {\rm e}^{i (\bg \cdot
 \br + q z^\prime)} \int d\br^\prime \int_{-t/2}^{t/2} dz^{\prime\prime} \nn\\
 & \times &
 \sum_\b \frac{\pd^2}{\pd r_\a  \pd r_\b} \frac {m_\b(\br^\prime)}
 {\sqrt{\left|\br - \br^\prime\right|^2 + \left(z^\prime -
z^{\prime\prime}\right)^2}}.
 \nn
 \eea

\noindent To calculate them, we express the lattice magnetization
$m_\b(\br^\prime)$ through its Fourier amplitudes:

 \[m_\b(\br^\prime) = \frac 1 {a^2} \sum_{\bg^\prime} {\rm e}^{-i \bg^\prime
 \cdot \br^\prime}  m_{\b,\bg^\prime},\]

 \noindent and then introduce the factor ${\rm e}^{i(\bg^\prime \cdot \br -
 qz^{\prime\prime})}$ into the integral in $d\br^\prime dz^{\prime}$,
 and the compensating factor ${\rm e}^{-i(\bg^\prime \cdot \br -
 qz^{\prime\prime})}$ into the integral in $d\br dz^{\prime\prime}$.
 Then the spatial integrations in $E_d$ are done accordingly to the formulas:

 \bea
&&\int_c d\br {\rm e}^{i (\bg - \bg^\prime) \cdot \br} = a^2
\d_{\bg,\bg^\prime},\nn\\
&&\,\quad\int_{-t/2}^{t/2} dz \int_{-t/2}^{t/2}
dz^{\prime\prime}{\rm e}^{i q (z^{ \prime
\prime} - z)} = \frac 4 {q^2} \sin^2 \frac {qt} 2,\nn\\
 && \qquad\quad \int_{-\infty}^{\infty} dz^\prime \int d\br^\prime {\rm
e}^{i\bg \cdot (\br - \br^\prime) + i q(z^\prime -
z^{\prime\prime})}\nn\\
&& \qquad\qquad \times\frac{\pd^2}{\pd r_\a \pd r_\b} \frac 1
{{\sqrt{\left| \br
- \br^\prime\right|^2 + \left(z^\prime - z^{\prime\prime}\right)^2}}}\nn\\
& & \qquad\qquad\qquad\qquad\qquad = \frac {4 \pi g_\a g_\b}{g^2 +
q^2}. \nn \eea

\noindent Finally, the momentum integration

 \[\int_{-\infty}^{\infty} \frac{\sin^2 (qt/2)} {q^2({g}^2 +
 q^2)} dq = \frac{\pi t}{2 {g}^2} f(g t)\]

\noindent leads to the result included in Eq. \ref{eq1}.

\end{document}